\documentclass[journal]{IEEEtran}
\usepackage{graphicx}
\usepackage{color}
\usepackage{paralist}
\usepackage{soul}
\usepackage{url}
\usepackage[boxed, linesnumbered]{algorithm2e}
\usepackage{epsfig}
\usepackage{amssymb}
\usepackage{amsmath}
\usepackage{amsfonts}
\usepackage{xspace}

\newcommand*{\ie}{\emph{i.e.}\@\xspace}
\newcommand*{\etal}{\emph{et al.}\@\xspace}

\makeatletter
\newcommand*{\etc}{%
    \@ifnextchar{.}%
        {\emph{etc}}%
        {\emph{etc.}\@\xspace}%
}

\begin{document}

\title{Modeling Multimodal Clues in a Hybrid Deep Learning Framework for Video Classification}
\author{Yu-Gang Jiang, Zuxuan Wu, Jinhui Tang, Zechao Li, Xiangyang Xue, Shih-Fu Chang \thanks{Y.-G. Jiang, Z. Wu and X. Xue are with the School of Computer Science, Fudan University, Shanghai, China. E-mail: \{ygj,zxwu,xyxue\}@fudan.edu.cn. J. Tang and Z. Li are with the School of Computer Science and Engineering, Nanjing University of Science and Technology, Nanjing, China. E-mail: \{jinhuitang,zechao.li\}@njust.edu.cn. S.-F. Chang is with the Department of Electrical Engineering, Columbia University, New York City, USA. E-mail: sfchang@ee.columbia.edu. }}

\markboth{Journal of \LaTeX\ Class Files,~Vol.~14, No.~8, August~2015}%
{Shell \MakeLowercase{\textit{et al.}}: Bare Demo of IEEEtran.cls for IEEE Journals}

\maketitle

\begin{abstract}
Videos are inherently multimodal. This paper studies the problem of how to fully exploit the abundant multimodal clues for improved video categorization. We introduce a hybrid deep learning framework that integrates useful clues from multiple modalities, including static spatial appearance information, motion patterns within a short time window, audio information as well as long-range temporal dynamics. More specifically, we utilize three Convolutional Neural Networks (CNNs) operating on appearance, motion and audio signals to extract their corresponding features. We then employ a feature fusion network to derive a unified representation with an aim to capture the relationships among features. Furthermore, to exploit the long-range temporal dynamics in videos, we apply two Long Short Term Memory networks with extracted appearance and motion features as inputs. Finally, we also propose to refine the prediction scores by leveraging contextual relationships among video semantics. The hybrid deep learning framework is able to exploit a comprehensive set of multimodal features for video classification. Through an extensive set of experiments, we demonstrate that (1) LSTM networks which model sequences in an explicitly recurrent manner are highly complementary with CNN models; (2) the feature fusion network which produces a fused representation through modeling feature relationships outperforms alternative fusion strategies; (3) the semantic context of video classes can help further refine the predictions for improved performance. Experimental results on two challenging benchmarks, the UCF-101 and the Columbia Consumer Videos (CCV), provide strong quantitative evidence that our framework achieves promising results: $93.1\%$ on the UCF-101 and $84.5\%$ on the CCV, outperforming competing methods with clear margins.
\end{abstract}


\IEEEpeerreviewmaketitle

\section{Introduction}
Classifying videos based on content semantics has been a hot research topic in multimedia for over a decade. Related techniques can be deployed in a multitude of applications such as video indexing, retrieval, advertising, \etc. The key enabling factors behind the significant technical progress in recent years are discriminative and robust feature representations that can not only withstand large intra-class variations but also effectively differentiate multiple classes. Some popular feature descriptors such as SIFT \cite{lowe2004distinctive} and HOG~\cite{dalal2005histograms} model spatial clues like texture, while others such as HOF~\cite{dalal2006} and trajectory features~\cite{wang2013action,jiang2015super,shi2016sequential}, focus on motion information, a fundamental nature of video depicting movements of objects among adjacent frames. Recently, deep neural networks, especially Convolutional Neural Networks (CNNs), have demonstrated great potentials for deriving robust features from raw data on a variety of tasks, including image classification~\cite{krizhevsky2012imagenet}, object detection~\cite{girshick2014rcnn}, speech recognition~\cite{DBLP:conf/icassp/GravesMH13}, \etc. Researchers have also attempted to apply deep learning techniques to the video domain. For instance, a straightforward extension is to stack multiple frames over time as inputs to CNNs for spatial-temporal feature learning~\cite{KarpathyCVPR14,DBLP:conf/icml/JiXYY10,Tran2015}. Different from these works, Simonyan \etal~\cite{DBLP:conf/nips/SimonyanZ14} disentangled video feature learning with two independent CNNs operating on RGB frames and stacked optical flow images to capture spatial and motion information, respectively. Final predictions are derived by linear combination of scores from the two CNNs and the results are competitive with state-of-the-art trajectory features~\cite{wang2013action}. 
However, these works merely focus on appearance and motion information in videos, ignoring the abundant long-range temporal clues therein because the training of CNNs totally neglects the order of inputs (\ie, RGB frames or stacked optical flow images). In addition, the motion CNN can only account for object movements within very short time periods. We believe this is not satisfactory for understanding video contents since different segments of videos usually correspond to different states of actions/events and their temporal order can assist recognition. For example, a ``celebrating birthday'' event could start with ``making a wish'', followed by ``blowing out candles'', and finally ends with ``eating cakes''. Moreover, audio signal is an indispensable component of video data, providing complementary clues to visual information. In the case of a ``celebrating birthday'' event, a birthday song is typically associated with the video.

Further, video semantics usually do not occur in isolation, and recognizing a class of interest could benefit from its semantic contextual relationships. For example, similar human motion patterns can be observed  in ``running'' and ``playing tennis'', and the likelihood of a video containing ``running'' could potentially help recognize ``playing tennis''. These useful clues are either overlooked or modeled with complicated models that are infeasible to scale up in most existing works. 



\begin{figure*}
\centering
\epsfig{file=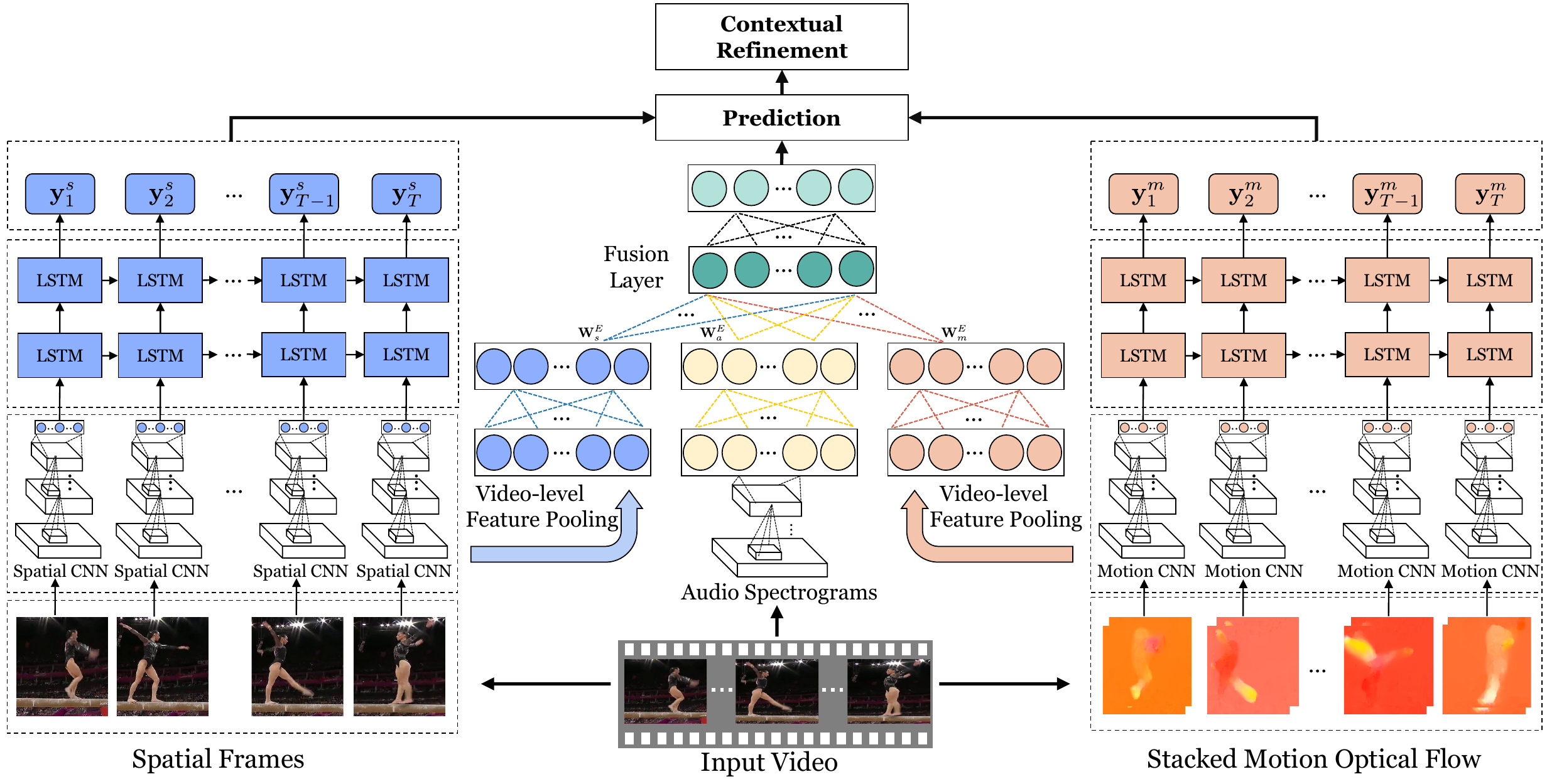, scale=0.71}
\caption{\label{fig:framework} The pipeline of the proposed hybrid deep learning framework. For a video clip, we first extract spatial, motion and audio features with three CNNs operating on video frames, stacked optical flow images and audio signals respectively. To capture long-range temporal dynamics in videos, we leverage two LSTM models with inputs of the extracted spatial and motion features. Further, we also utilize a feature fusion network to integrate multiple features into a unified representation to perform classification with carefully designed regularizations aiming to exploit feature relationships. Finally, we combine the outputs of the LSTM models with feature fusion network with refinement to generate final prediction scores. See texts for more discussions.}
\end{figure*}

To mitigate these limitations, we propose a hybrid deep learning framework for video categorization that is designed to explore the abundant multimodal clues embedded in videos, including static spatial, motion patterns, audio information and long-range temporal coherence as well as the contextual relationships among video semantics. Motivated by the great success of Recurrent Neural Networks (RNNs) for sequence modeling tasks~\cite{DBLP:conf/icassp/GravesMH13,graves2005framewise}, we leverage Long Short Term Memory (LSTM), a variant of RNNs with memory units and different functional gates, to account for temporal information. Furthermore, different from existing methods that integrate features in a straightforward and heuristic way by either feature concatenation or score averaging, we are interested in exploring the feature correlations. To this end, we apply a deep neural network with carefully designed regularizations~\cite{TPAMI-fcvid} to integrate the extracted static appearance, short-term motion and audio features. Then we combine the predictions from this network with the outputs of LSTMs. Finally, we refine the prediction scores in consideration of contextual relationships among video semantics in a simple yet effective manner.


The framework is illustrated in Figure~\ref{fig:framework}. In particular, we first compute spatial appearance, short-term motion (based on stacked optical flow images) and audio features with CNN models. The spatial and motion features are further utilized as inputs of LSTMs to capture the long-range temporal temporal clues. Then, a feature fusion network takes the video-level features (spatial, motion and audio) to derive a unified representation for predicting video semantics. The outputs of the feature fusion networks are further combined with scores from the LSTMs and then refined by taking advantage of the contextual relationships of video semantics. The main contributions are summarized as follows:

\begin{itemize}
\item We fully exploit a variety of multimodal clues in a hybrid deep learning framework for improved video categorization, including static spatial appearance, motion and audio information, long-range temporal coherence and contextual relationships among video semantics.
\item We demonstrate that the LSTMs, modeling the long-range temporal information in video sequences through an explicitly recurrent manner, are highly complementary with CNNs.

\item We resort to the rich contextual relationships among video semantics in a simple yet effective way to further refine predictions for improved performance.

\item We conduct experiments on two challenging benchmarks, and the experimental results provide strong quantitative evidence that our framework achieves promising results, outperforming competing methods with clear margins.
\end{itemize}

This work extends from a conference paper~\cite{Wu2015} by incorporating audio and semantic contextual relationships in the hybrid framework. New experiments are conducted to verify the effectiveness of the technical extensions and extra amplified discussions are provided throughout the paper. The remaining sections are organized as follows. We first review related works in Section~\ref{sec:related} and elaborate the proposed hybrid deep learning framework in Section~\ref{sec:method}. We then present and discuss the experimental results and comparisons in Section~\ref{sec:exp}. Finally, Section~\ref{sec:conclusion} concludes this paper.

\section{Related Works}
\label{sec:related}
We divide the discussions of related works into the following five subsections. 
\subsection{Hand-crafted Features}
There is a large body of literature on video classification in the multimedia community (see~\cite{IJMIR:EventSurvey} for a survey). Among these works, designing powerful feature representations is an important topic due to the significant role of features in a typical video recognition pipeline. The success of image descriptors like SIFT and HoG has spurred the developments of video representations by considering the temporal feature of videos. For example, Harris corner detector is extended into 3D volumes to identify space-time interest points~\cite{laptevSTIP}. Similarly, based on HoG features, 3D spatial-temporal gradients are derived as local descriptors for action recognition~\cite{klaser2008spatio}. Wang \etal proposed to track densely sampled local patches over time in an optical flow field to compute dense trajectory features, which achieved superior performance on a variety of benchmarks when coupled with quantization techniques like Bag-of-Words and Fisher Vector~\cite{oneata2013action,han2017vrfp}. However, these video representations focus on modeling local motion patterns within short time periods and the feature encoding methods while powerful totally discards the temporal information of videos. 

\subsection{CNN Representations}
Different from hand-crafted features, recent advances on CNNs in image~\cite{krizhevsky2012imagenet,girshick2014rcnn} and speech domain~\cite{DBLP:conf/icassp/GravesMH13} have encouraged works to learn features directly from raw video data. The most straightforward way to utilize CNN on video data is stacking frames as inputs with an aim to learn spatial-temporal features using 3D convolutions~\cite{DBLP:conf/icml/JiXYY10,KarpathyCVPR14,Tran2015}. However, these works demonstrate worse performance than state-of-the-art trajectory features~\cite{wang2013action}. This might result from the difficulty to learn 3D features with insufficient training data. To effectively model 3D signals, Simonyan \etal proposed to utilize two independent CNNs to capture spatial and motion information operating on RGB frames and stacked optical flow images, separately. Based on this approach, Wang \etal proposed to learn the transformation between two states triggered by actions~\cite{Wang2016}. Feichtenhofer \etal experimented with different fusion approach to combine spatial and temporal features~\cite{Feichtenhofer2016}. During the training process of CNNs, the temporal order of frames and stacked optical flow images is discarded and thus the temporal structures of videos are ignored.

\subsection{Temporal Information}
Graphical models, including Conditional Random Fields (CRF), Hidden Markov Models (HMM), \etc, have been widely adopted to capture long-term temporal structures~\cite{vail2007conditional,wang2009max,tang2012learning}. For example, Tang \etal proposed a variable duration HMM to model state changes in videos~\cite{tang2012learning}. Instead of using graphical models, Fernando \etal utilized a ranking machine to account for the temporal order of frames. Wang \etal proposed the temporal segment networks, which used a consensus function to combine segment scores generated by two-stream networks~\cite{Wang2016a}. 

Many works resort to LSTM to capture temporal dynamics in videos due to its great success in sequential modeling tasks like speech recognition~\cite{DBLP:conf/icassp/GravesMH13} and video captioning~\cite{Sutskever:2014ty}. Srivastava \etal proposed to learn video features using an auto-encoder framework~\cite{Srivastava2015} based on LSTMs. Donahue \etal utilized two LSTM models using spatial and motion features extracted from CNN models~\cite{Donahue2015}. Ng \etal further deepened LSTM to five layers and experimented with several pooling strategies~\cite{Ng2015}. Our work leverages LSTMs for temporal modeling to explicitly complement the limitation of the frame-based CNN models.

\subsection{Feature Fusion}
Extensive works have been conducted on the fusion of multiple features, the complementarity of which is expected to promote classification accuracy. There are two popular fusion strategies, \ie early fusion and late fusion performed at the feature level and the classification score level, respectively~\cite{snoek2005early,yang2013multi}. Typically, early fusion integrates features by direct concatenation~\cite{wang2013action} or linear combination of their kernels~\cite{zhang2007local} before classification.
In addition, Multiple Kernel Learning (MKL) can also be applied to combine feature kernels, where the weights are automatically learned. Late fusion, on the other hand, combines prediction scores from multiple classifiers, each of which is independently trained with a single feature~\cite{ye2012robust,liu2013sample}. Both fusion methods are popular due to their simplicity, however, they assume the features or prediction scores are explicitly complementary to one another and fail to consider potential hidden correlations among features. Recently, Srivastava \etal utilized Deep Boltzmann Machines (DBM) to derive an embedding of images and texts~\cite{Srivastava2015} and Ngiam \etal used deep auto-encoder to learn the relationships between different modalities~\cite{ngiam2011multimodal}. Wu \etal proposed to explore feature and class relationships~\cite{mm14:videoclassification} by imposing trace norms. In this work, to alleviate computational complexity, we adopt a regularized neural network to automatically learn dimension-wise correlations of features extracted from state-of-the-art CNN models.

\subsection{Contextual Relationships}
As aforementioned, the co-occurrence of video semantics, serving as context, can provide useful information. For example, Rabinovich~\etal proposed to incorporate the semantics context information with a CRF model~\cite{rabinovich2007objects}. Jiang \etal modeled the class relationships with a semantic diffusion algorithm~\cite{iccv09:dasd}. Deng \etal leveraged a graphical model to encode label hierarchies for improved image classification performance~\cite{deng2009imagenet}. Wu \etal proposed to capture the relationships of video semantics by regularizing the classification process~\cite{mm14:videoclassification}. Chen \etal utilized confusion matrix to predict the context of a category when training CNNs~\cite{chen2015webly}. In our paper, we propose to utilize confusion matrix as contextual relationships derived from trained models, to refine the prediction scores as a post-processing step. Therefore, the recognition of a class of interest can benefit from its related classes.

\section{Methodology}
\label{sec:method}
We now elaborate the proposed hybrid deep learning framework illustrated in Figure~\ref{fig:framework}. We first introduce the multimodal features extracted by CNN models, and present the modeling the temporal dynamics in videos with LSTM models. Then we describe the feature fusion framework which is designed to model feature correlations. Finally, we introduce the contextual refinement.

\subsection{Spatial, Motion and Audio CNN Features}
CNN models usually contain alternating convolutional and pooling layers to learn features from input images, followed by fully-connected (FC) layers for classification. In our framework, we first compute spatial and motion features based upon the two-stream approach~\cite{DBLP:conf/nips/SimonyanZ14}, where two independent CNNs are trained with RGB frames and stacked optical flow images, respectively. More concretely, the spatial stream models static appearance information like texture  from sampled video frames as in conventional CNNs for image classification. The motion CNN takes stacked optical flow images as inputs to capture object movements within a short time window. Optical flow is an explicit form of motion patterns derived by computing displacement vector fields between two adjacent frames, whose horizontal and vertical components are then used to generate two images. Multiple optical flow images are further stacked to represent motion information in a short period, upon which convolution is performed. Given a video at testing phase, each stream averages prediction scores produced by soft-max layer from 25 uniformly sampled frames (or stacked optical flow images) and the scores from the two streams are further linearly combined as the final prediction. In our work, we compute the outputs from the first FC layer of two CNNs, which are observed to be effective in many tasks~\cite{Razavian2014}, as the spatial and motion features to model long-term temporal structures and explore their correlations for improved performance. 

In addition, we also utilize a CNN model to capture the acoustic information in videos as a compliment to visual information. Particularly, we convert the 1D soundtrack extracted from a video clip into a 2D spectrogram image with Short-Time Fourier Transformation, demonstrating changes of frequency-scale along with time. Then, inspired by~\cite{VandenOord2013}, we take the spectrograms as inputs to a CNN network to capture the acoustic clues.

\subsection{Temporal Modeling with LSTM}
As aforementioned, the two-stream approach focuses only on appearance and short-time motion information, which ignores the long-term temporal dynamics in videos. Therefore, we employ the LSTM model due to its great success in sequential modeling tasks~\cite{DBLP:conf/icassp/GravesMH13,Donahue2015,Yao2015}. Compared with conventional RNN models that map input data recursively to outputs through hidden states, an LSTM additionally incorporates a memory cell with multiple gates governing information into and out of the cell, enabling it to model long sequences without suffering from the ``vanishing gradients'' effect.



\begin{figure}[t]
\centering
\epsfig{file=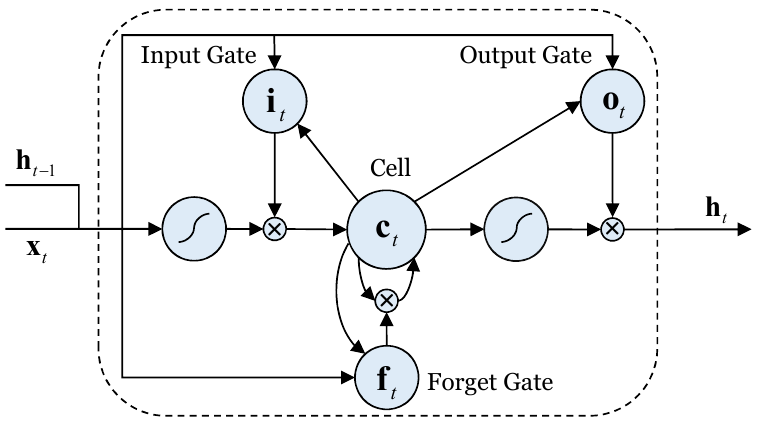, scale=1}
\caption{\label{fig:lstmunit}An illustration of an LSTM unit.}
\end{figure}

Formally, an LSTM takes a sequence $({\bf x}_1,{\bf x}_2,\ldots,{\bf x}_T)$ as inputs and maps it to an output sequence  $({\bf y}_1,{\bf y}_2,\ldots,{\bf y}_T)$ by recursively computing activations of the units from $t=1$ to $t = T$ as following:
\begin{align*} 
& {\bf i}_t=\sigma({\bf W}_{xi}{\bf x}_t+{\bf W}_{hi}{\bf h}_{t-1}+{\bf W}_{ci}{\bf c}_{t-1}+{\bf b}_i), \\ 
& {\bf f}_t=\sigma({\bf W}_{xf}{\bf x}_t+{\bf W}_{hf}{\bf h}_{t-1}+{\bf W}_{cf}{\bf c}_{t-1}+{\bf b}_f), \\
& {\bf c}_t={\bf f}_t{\bf c}_{t-1}+{\bf i}_t\tanh({\bf W}_{xc}{\bf x}_t+{\bf W}_{hc}{\bf h}_{t-1}+{\bf b}_c), \\
& {\bf o}_t=\sigma({\bf W}_{xo}{\bf x}_t+{\bf W}_{ho}{\bf h}_{t-1}+{\bf W}_{co}{\bf c}_{t}+{\bf b_o}), \\
& {\bf h}_t={\bf o}_t\tanh({\bf c}_t).
\end{align*}
Here, at the $t$-th time step, we denote the input features as ${\bf x}_t$ and the hidden states as ${\bf h}_t$. And ${\bf c}_t$ represents the contents of the memory unit. The activations of the input, forget and output gates are represented as ${\bf i}_t, {\bf f}_t, {\bf o}_t$, respectively.
${\bf W}_{\alpha\beta}$ represents the transition weights from component $\alpha$ to component $\beta$, and ${\bf b}_\alpha$ is the corresponding bias term. In addition, $\sigma(x) = \frac{1}{1+e^{-x}}$ is the non-linear sigmoid function. We present the structure of an LSTM unit in Figure~\ref{fig:lstmunit}.

The memory cell regulated by different non-linear gates enables LSTM model to store information progressively. More concretely, for the $t$-th time step, the current feature representation ${\bf x}_t$ together with information from the past ${\bf h}_{t-1}$ are fed into all gates and the memory cell. Past information stored in the memory cell ${\bf c}_{t-1}$ regulated by the activations of the forget gate ${\bf f}_t$ is linearly combined with the squashed inputs multiplied by the activation of the input gate ${\bf i}_t$ to generate the current ``memory''. This facilitates the LSTM model to learn when to utilize current information or forget previous contents. Furthermore, the information that will be used for future states is regulated by the output gate ${\bf o}_t$. The interactions between the memory units and these multiplicative gates allow LSTM to capture
the temporal dynamics in long sequences, making it a natural fit for video classification.

One can also stack hidden states to deepen the LSTM model aiming to increase its discriminative power. A softmax layer can then be applied on top of the hidden states to obtain the prediction scores at each time-step. The training of LSTM is usually conducted with stochastic gradient descent using the Back-Propagation Through Time (BPTT) algorithm~\cite{graves2005framewise}.

In our framework (illustrated in Figure~\ref{fig:framework}), we model the temporal information in videos with two LSTMs, operating on a spatial feature sequence $({\bf x}_1^s,{\bf x}_2^s,\ldots,{\bf x}_T^s)$ and a motion feature sequence $({\bf x}_1^m,{\bf x}_2^m,\ldots,{\bf x}_T^m)$, respectively. Once the model is trained, the two LSTM models will produce two sets of predictions: $({\bf y}_1^s,{\bf y}_2^s,\ldots,{\bf y}_T^s)$ for the spatial stream and $({\bf y}_1^m,{\bf y}_2^m,\ldots,{\bf y}_T^m)$ for the motion stream. We compute the prediction from the last time step ${\bf y}_T$ of a sequence as the score for the entire video, because it contains information from all previous steps.

\subsection{Regularized Feature Fusion Network}
The spatial, motion and audio features characterize the same video from different perspectives (\ie, person-related static appearance information, body motions and sound), and thus certain correlations between these features might exist. We posit that an ideal unified representation is expected contain information shared by multiple features as well as the special aspect of each feature. This requires modeling feature relationships explicitly instead of uniform fusion approaches. To this end, we utilize a regularized feature fusion network~\cite{TPAMI-fcvid} to fully exploit feature relationships (see Figure~\ref{fig:framework}). Given a video clip, we compute its video-level appearance and motion features by simply averaging descriptors from all frames. The spatial, motion and audio features are separately transformed into a higher space with one hidden layer. We then apply one hidden layer to absorb all the features to derive a unified representation, regularized by carefully designed norms to explore feature relationships.

We represent the $n$-th training video as a 4-tuple $({\bf x}^s_n, {\bf x}^m_n, {\bf x}^a_n, {\bf y}_n)$, where ${\bf x}^s_n = \sum_{t=1}^T {{\bf x}^s_{n,t}}\in \mathbb{R}^{d_s}$ and ${\bf x}^m_n = \sum_{t=1}^T {{\bf x}^m_{n,t}} \in \mathbb{R}^{d_m}$ denote the video-level spatial and motion descriptors respectively, ${\bf x}^a_n \in \mathbb{R}^{d_a}$ as the audio feature derived from the audio CNN and ${\bf y}_n$ is the corresponding ground-truth label. We first consider the training of a neural network with a single feature as inputs. Let $g(\cdot)$ denote the non-linear function approximated by the neural network. To learn the optimal weights of model, we minimize the following objective function:
\begin{equation}
\label{eq:nn}
    \min_{{\bf W}} ~\sum_{i=1}^{N} \|  g({\bf x}_{i})-{\bf y}_i \|^2 + \lambda_1 \Phi({\bf W}).
\end{equation}
Here $N$ denotes the number of videos in the training set and the first item is the empirical loss, and the second term is a penalty on the weight matrices to prevent over-fitting or forcing sparsity, depending on different choices of norms. 

We now introduce the fusion of multiple features in a regularized framework. Given three types of features, we first perform feature transformation independently and then integrate them to derive a fused representation. In the fusion process, we impose a structural $\ell_{21}$ norm to explore the relations of the features. The optimization problem now becomes:

\begin{equation}
\label{eq:obj1}
  \min_{{\bf W}}~\mathcal{L} + \lambda_1 \Phi({\bf W})
+\frac{\lambda_{2}}{2}\left \| {\bf W}^{E} \right \|_{2,1}.
\end{equation}
Here $\mathcal{L} =  \sum_{i=1}^{N} \|  g({\bf x}_{i}^{s},{\bf x}_{i}^{m}, {\bf x}_{i}^{a})-{\bf y}_i \|^2$,  ${\bf W}^E = [{\bf W}^{E}_{s},  {\bf W}^{E}_{a}, {\bf W}^{E}_{m}] \in \mathbb{R}^{P \times D}$ represents the stacked weights for the $E$-th layer, where $D = d_s + d_m + d_a$ and $P$ denotes dimension for the unified feature representation. 

Compared with Equation~(\ref{eq:nn}), an $\ell_{21}$ norm is appended to regularize the fusion process of the the $E$-th layer aiming to exploit feature relationships. The $\|{\bf W}\|_{2,1}$ is defined as $\sum_i \sqrt{\sum_j w_{ij}^2}$, and we can see that it first computes $\ell_2$ norm for each row (weights of the three features), and then $\ell_1$ norm for the resulting vector, which will force the matrix ${\bf W}^E$ to be row sparse and produce similar zero/nonzero patterns for the columns. 
In other words, the norm will be minimized when there are only a few non-zero rows in the weight matrix, which serve as the shared discriminative information of these features.

As aforementioned, we posit a good unified representation should be derived without loss of information of original features, which requires the fusion process not only leverages feature correlations but also preserves the special information of each feature. As such, we additionally regularize the fusion process with an $\ell_1$ norm and rewrite Equation~(\ref{eq:obj1}) as following:
\begin{equation}
\label{eq:obj}
  \min_{{\bf W}}~\mathcal{L} + \lambda_1 \Phi({\bf W}) +\frac{\lambda_{2}}{2}\left \| {\bf W}^{E} \right \|_{2,1} + \lambda_{3}\left \| {\bf W}^{E} \right \|_{1,1}.
\end{equation}
The regularizer $\|{\bf W}^E\|_{1,1}$ complements the $\|{\bf W}^E\|_{2,1}$ norm to be robust by preventing it from sharing incorrect information, which enables different features to select different neurons (\ie, the unique information of these features).

We now move on to discuss the optimization in Equation~(\ref{eq:obj}), which is nonconvex because of the multi-layer neural network. Therefore, we train the network using back-propagation with gradient descent method in two scenarios:

\IncMargin{0.66em}
\begin{algorithm}
\SetKwData{Left}{left}\SetKwData{This}{this}\SetKwData{Up}{up}
\SetKwFunction{Union}{Union}\SetKwFunction{FindCompress}{FindCompress}
\SetKwInOut{Input}{Input}\SetKwInOut{Output}{Output}
\Input{$\mathbf{x}_{n}^s$, $\mathbf{x}_{n}^m$ and $\mathbf{x}_{n}^a$: the video-level spatial, motion and audio CNN features of the $n$-th video\;
~~~~~~~~~~~$\mathbf{y}_n$: the corresponding ground-truth label\;
~~~~~~~~~~~randomly initialized weights ${\bf W}$;}
\Begin{
\For{$epoch\leftarrow 1$ \KwTo $M$}{
 Run a feed-forward pass through the network to obtain perdition error\;
 \For{$l \leftarrow L$ \KwTo 1} {
   Gradient descent with Eqn.~(\ref{eq:update})\;
   \If{~$l==E$~}{
   Update the weights with proximal operation with Eqn.~(\ref{eq:sol})\;
   }}
}}
\caption{Algorithm for training the regularized feature fusion network.}\label{alg}
\end{algorithm}\DecMargin{0em}

\begin{enumerate}
    \item The $E$-th layer. Since the regularization is imposed only on the $E$-th layer, we treat it differently when performing gradient descent. The difficulty of the optimization here lies in the last two non-smooth terms, which are non-differentiable. Thus we cannot directly apply gradient descent. Instead, we utilize the proximal operation to evaluate their gradients. More specifically, we split the objective function into two components:
        \begin{align*}
            p &= \mathcal{L} +  \lambda_1 \Phi({\bf W}), \\
            q &= \frac{\lambda_{2}}{2}\left \| {\bf W}^{E} \right \|_{2,1}+\lambda_{3}\left \| {\bf W}^{E} \right \|_{1,1}.
        \end{align*}
        Here $p$ is a smooth function whose gradients are easy to obtain and $q$ is a non-smooth function. We utilize a proximal operator to update the weights for the $i$-th iteration:
        \begin{equation*}
            ({\bf W}^E)^{(i)} = \text{Prox}_q (({\bf W}^E)^{(i)} - \nabla p(({\bf W}^E)^{(i)})),
        \end{equation*}
        where $\text{Prox}_q({\bf W}) = \arg\min_{\bf V} \|{\bf W}-{\bf V}\| + q(V)$. Note that $q$ here is a combination of $\ell_{21}$ and $\ell_{11}$ norms, and thus the proximal operator can be can be derived as:
        \begin{equation}
        \label{eq:sol}
            {\bf W}^E_{r\cdot} = \left( 1- \frac{\lambda_2}{\|{\bf U}_{r\cdot}\|_2} \right) {\bf U}_{r\cdot}, \forall r = 1,\cdots, P,
        \end{equation}
        where ${\bf U}_{r\cdot} = \max\{|{\bf V}_{r\cdot}| - \lambda_3,0 \} \cdot sign[{\bf V}_{r\cdot}]$, and ${\bf W}_{r\cdot},{\bf U}_{r\cdot},{\bf V}_{r\cdot}$ represents the $r$-th row of matrix ${\bf W}, {\bf U}$ and ${\bf V}$, respectively.
    \item  Other layers. Since there are no non-smooth regularizations for other layers, we compute their gradients directly and then update the weight matrix with gradient descent as in \cite{bengio2012practical}. Let ${\bf G}^l$ represent the gradients of ${\bf W}^l$, the weight matrix of the $l$th layer is updated as:
    \begin{equation}
    \label{eq:update}
    {\bf W}^l = {\bf W}^l - \eta {\bf G}^l.
    \end{equation}
\end{enumerate}

Although the two regularization norms in function $q$ incur extra computation cost, it is worth noting that the complexity of computing the proximal operator is $O(P \times D)$, which is fast to evaluate. The proposed method is also general for fusing more features at a linearly growing computational cost rather than cubic cost as in \cite{TPAMI-fcvid}. The overall training process of the feature fusion framework is presented in Alg.~\ref{alg}.



\subsection{Contextual Relationships}
Given the classification scores from the two LSTMs and the regularized feature fusion network, accounting for spatial, motion, audio and long-term temporal clues in videos, we are interested in incorporating contextual relationships to further refine the outputs for improved performance. More specifically, for each video sample, we first linearly average the probabilities to obtain a compact prediction. Then we utilize a simple approach to refine the prediction with contextual relationships, which provide useful information of semantics co-occurrence. For example, ``baseball'' is more related to ``soccer'' than ``diving'', since ``diving'' contains totally different motion patterns. And if the likelihood for the video to be ``soccer'' is extremely low, then it is also unlikely to be ``baseball''. Existing works often resort to external knowledge like WordNet or word vectors to obtain class relationships, which are either hand-crafted or trained on text corpus and hence fail to consider visual patterns. In our work, we simply rely on the trained models to produce class relationships by computing the confusion matrix, which is a good indicator on how classes are related. 

Formally, for a total of $C$ classes, we denote $f(\cdot) \in \mathbb{R}^{C}$ as the mapping from the input to the linearly averaged prediction and then the confusion matrix ${\bf R} \in \mathbb{R}^{C \times C}$ is defined as following:
\begin{align}
{\bf R}_{ij}  = \frac{1}{|C_i|}\left| \left\{ ({\bf x}, {C_i})  \in \mathcal{V}: \arg\max f({\bf x)} = C_j  \right\} \right|.
\label{eq:confusionmatrix}
\end{align} 
Here, $\mathcal{V}$ is the validation set and $\left| \cdot \right|$ is the cardinality function. When $i \neq j$, ${\bf R}_{ij}$ measures 
the number of samples originally belongs to the $C_i$ class but are misclassified into $C_j$. It is easy to understand that if $C_i$ and $C_j$ are close, the value ${\bf R}_{ij}$ will be large since they are difficult to separate. Then for the $i$-th video sample, we refine its prediction score by:
\begin{align}
{\bf p}_i = {\bf R}f({\bf x}_i),
\label{eq:refine}
\end{align}
where ${\bf p}_i$ is the final probability for the $i$-th video sample. The recognition of a class of interest can benefit from the contextual relationships in that information from its related classes is utilized to adjust its confidence based on semantic co-occurrence. {\color{black} Note that researchers also employ multi-label loss functions like hinge loss or ranking loss~\cite{Chatfield14} to consider context in an explicit way but they are not suitable for single-label recognition tasks. Our approach models contextual relationships among classes by analyzing their appearance and motion patterns, and thus it is general to both multi-label and single-label scenarios.}

\subsection{Discussion}
The proposed framework is able to model a comprehensive set of multimodal features, including static appearance, motion patterns in a short time window, long-range temporal dynamics and acoustic clues, which are all critical for understanding video contents since they describe videos from different perspectives. In our framework, we train different components independently rather than jointly in an end-to-end manner. Although training jointly is theoretically feasible, it would require extra training samples to prevent under-fitting in the complicated process and it is observed in~\cite{Donahue2015} the performance gain of joint training is rather marginal. In addition, separate training ensures flexibility in the framework, since a component can be replaced easily without incurring the re-training of the whole complex framework. For example, one can easily update the framework with more powerful CNN models like GoogleNet~\cite{Szegedy2015} and ResNet~\cite{He2016a} or better RNN models~\cite{DBLP:journals/corr/ChungGCB15}. The main purpose of this paper is to demonstrate that a comprehensive set of features are demanded for improved video classification. In addition, in this work, we mainly demonstrate audio information captured by a CNN model can serve as an effective complement to visual information, and thus we do not investigate modeling temporal audio dynamics with LSTMs.

\section{Experiments}
\label{sec:exp}
In this section, we first introduce the experimental settings and then discuss the results of the proposed hybrid deep learning framework on two popular benchmarks.
\subsection{Experimental Setup}
\subsubsection{Datasets}
To investigate the effectiveness of the proposed hybrid deep learning framework, we utilize the following two benchmarks:
\begin{itemize} 

\item{UCF-101}~\cite{ucf101}. The UCF-101 benchmark is a widely adopted dataset for human action recognition, which contains 13,320 video clips manually annotated into 101 human actions, totaling 27 hours. We conduct experiments using three training and testing splits following the protocol defined in~\cite{THUMOS14}. Performance is measured by the average classification accuracy of all three splits.

\item{Columbia Consumer Videos (CCV)}~\cite{icmr11:consumervideo}. It consists of 9,317 videos collected from YouTube belonging to 20 categories, including ``basketball'', ``wedding dance'', ``soccer'', \etc.  Following~\cite{icmr11:consumervideo}, we utilize a training set of 4,659 videos and a testing set of 4,658 videos. We compute average precision for each class and report the mean AP over all classes.

\end{itemize}

\subsubsection{Implementation Details}
We utilize the VGG\_19 network~\cite{Simonyan2015} to extract spatial features and the CNN\_M model~\cite{DBLP:conf/nips/SimonyanZ14} to compute motion and audio features, due to their expressive performance on the ImageNet ILSVRC-2012 validation set: a 7.5\% and 13.5\% top-5 error rates, respectively. We first pre-train the spatial and audio CNN with ImageNet data and then fine-tune the network on video frames and spectrograms respectively. Note that for the audio CNN we observe better performance with pre-training though the images are spectrograms. Due to the lack of existing models trained on 20 channels (the input data format for the motion CNN), we train the motion CNN from scratch. To further promote the performance, we also employ simple data augmentation methods like cropping and flipping as in~\cite{DBLP:conf/nips/SimonyanZ14}.

We apply stochastic gradient descent using back-propagation to train the CNN models. We adopt a batch size of 256 and fix the momentum to be 0.9. To fine-tune the spatial and audio CNN, we first set the initial learning rate to $10^{-3}$ and decay it by a factor of 10 after every 14K iterations. Different from~\cite{DBLP:conf/nips/SimonyanZ14}, we begin with a smaller rate rather than $10^{-2}$. To train the motion network, we set the initial learning rate to $10^{-2}$, and then decay it by a factor of 10 after every 100K iterations. We adopt the popular Caffe~\cite{jia2014caffe} toolbox with modifications to support parallel training on multiple GPUs for implementations.

To capture the long-range temporal dynamics, we utilize two two-layer LSTMs operating on spatial and motion CNN features respectively. Both LSTMs contain 1,024 hidden neurons for the first layer and 512 units for the second layer. We train the network with a parallel implementation of Back-Propagation Through Time (BPTT) algorithm. The mini-batch size is set to 10 and the maximal iterations to be 150K. We also fix the learning rate and momentum to $10^{-4}$ and 0.9 respectively.


Finally, to learn the optimal weights for the feature fusion network, we follow the procedures described in Alg.~\ref{alg}. The network contains four hidden layers shown in Fig~\ref{fig:framework}. More concretely, we first employ a layer with 200 neurons for each of the spatial, motion and audio feature for independent feature transformation, followed by one layer with 200 neurons to perform feature fusion. The  derived unified feature representations are further trained to categorize videos into semantic classes. We utilize a learning rate of 0.7 and fix $\lambda_{1}$ to $3 \times 10^{-5} $ to prevent over-fitting. $\lambda_{2}$ and $\lambda_{3}$ are selected using cross-validation.

\subsubsection{Compared Approaches}

To evaluate the proposed framework, we compare with the following alternative competing methods: (1) \textbf{Spatial CNN}, \textbf{Motion CNN} and \textbf{Audio CNN}, which are independently trained with raw RGB frames, stacked optical flow images and audio spectrograms; (2) \textbf{Spatial LSTM} and \textbf{Motion LSTM}, which denote LSTM models operating on extracted spatial and motion CNN features respectively; (3) \textbf{SVM-based Early Fusion (SVM-EF)}, which averages three $\chi^2$-kernels derived from spatial, motion and audio features for classification with an SVM; (4) \textbf{SVM-based Late Fusion (SVM-LF)}, which employs a separate SVM for each feature and then linearly average their prediction scores; (5) \textbf{Multiple Kernel Learning (SVM-MKL)}, which integrates three features using the $\ell_p$-norm MKL~\cite{kloft2011lp} with $p=2$; 
(6) \textbf{Early Fusion with Neural Networks (NN-EF)}, which performs classification with a 4-layer neural network operating on the concatenated features; (7) \textbf{Late Fusion with Neural Networks (NN-LF)}, which combines predictions from three individual neural networks trained on three types of features respectively; (8) \textbf{Multimodal Deep Boltzmann Machines (M-DBM)}~\cite{ngiam2011multimodal,srivastava2012multimodal}, which performs feature fusion in a DBM without regularizations; (9) \textbf{RDNN} \cite{mm14:videoclassification}, which utilizes a different regularization scheme with higher computational complexity.

Notice that the first two classes of methods are components of the proposed framework and we report their performance independently to better analyze their contribution in the overall framework. The remaining seven methods aim to integrate the spatial, motion and audio features to improve classification performance.

\subsection{Results and Discussions}
\subsubsection{Multimodal Representations}
\paragraph{Temporal Modeling} In this section, we investigate the effectiveness of LSTMs on modeling the long-range temporal dynamics in video sequences. Table~\ref{tbl:lstm} presents the results of different methods on UCF-101 and CCV. We first compare the performance of LSTM models with CNNs as shown in the top two groups. Since CNN models fail to take the temporal order of frames into consideration, we expect the performance of CNN models is worse than LSTMs. On UCF-101, we can see that Spatial LSTM slightly outperforms Spatial CNN, but the Motion LSTM is marginally worse than Motion CNN. Since the motion LSTM takes stacked optical flow images as inputs, we posit this might result from the lack of training data to learn the optimal weights unlike the training of Spatial LSTM, where a large number of redundant frames could be utilized.

For CCV, CNN models perform consistently better than LSTM models on both spatial and motion streams. Compared to UCF-101, CCV contains more diversified and noisy videos without post-editing, whose duration are also significantly longer than those in UCF-101 (in average, 80 seconds \textit{vs.} 8 seconds). Therefore, the noises in such videos could significantly degrade the performance of LSTM models. The noisy nature of CCV videos can also be reflected by the relatively low performance of motion streams operating on optical flow images, which are sensitive to camera motions and cluttered backgrounds.

\begin{table}[h]
\begin{center}
\caption{Performance of the LSTM and the CNN models on UCF-101 and CCV. ``+" indicates model fusion, which simply uses the average prediction scores of different models.}
\begin{tabular}{|c||c|c|}
\hline
                  & UCF-101       & CCV     \\ \hline   \hline          
Spatial CNN           & 80.1\%      & 75.0\%  \\ 
Spatial LSTM          & 83.3\%      & 43.3\%  \\ \hline   \hline
Motion CNN            & 77.5\%      & 58.9\%      \\ 
Motion LSTM           & 76.6\%            & 54.7\%      \\ \hline   \hline

Audio CNN 			 & 16.2\%				& 21.5\% \\ \hline \hline
CNN + LSTM (Spatial)    & 84.0\%            & 77.9\%     \\ \
CNN + LSTM (Motion)       & 81.4\%      & 70.9\%     \\ 
CNN + LSTM (Spatial \& Motion)               
              &  90.1\%            &  81.7\%    \\ \hline \hline
CNN + LSTM (Spatial \& Motion) + Audio & 90.3\% & 82.4\% \\ \hline
\end{tabular}
\label{tbl:lstm} 
\end{center}
\end{table}

\paragraph{Audio Modeling} The performance of audio CNN is presented in the middle of Table~\ref{tbl:lstm}. Audio CNN operating on spectrograms achieves 16.2\% and 21.5\% on UCF-101 and CCV respectively. Note that the performance on UCF-101 is measured by mean accuracy over 101 classes, however only 51 categories contain soundtracks and thus the actual accuracy is 32.1\%. Audio signals are usually not robust and discriminative as visual clues due to the noises in video backgrounds.

\paragraph{Feature Complementarity} We now study whether the extracted multimodal representations are complementary through linearly averaging the outputs of the trained models. Here we only adopt simple late fusion and we will experiment with different fusion strategies in Sec.~4.2.2.

Results are summarized in the bottom two groups of Table~\ref{tbl:lstm}. We first combine CNN and LSTM models for both spatial and motion streams, and the fusion offers significant performance gains on both benchmarks. The combination of CNN and LSTM on the spatial stream offers 0.7\% and 2.9\% improvements over the best single model on UCF-101 and CCV, respectively. On the motion stream, the performance gains of fusion are more noticeable, 3.9\% and 12\% on UCF-101 and CCV. The consistent trend when fusing CNN with LSTM models on both streams confirms the complementarity of these features. Further, we also combine all spatial and motion models, offering 90.1\% and 81.7\% on UCF-101 and CCV respectively. This clearly verifies that spatial and motion features are very complementary. In addition, we also incorporate audio clues to complement the visual information, and this entire set of features attains the highest performance on both datasets: 90.3\% and 82.4\%. Therefore, we believe a successful video classification system should integrate all these features. 

\begin{table}[t]
\begin{center}
\caption{\label{tbl:fusion} Performance comparison on UCF-101 and CCV, using various fusion approaches to combine the multimodal clues. }
\begin{tabular}{|c||c|c|}
\hline
            & UCF-101      & CCV                              \\ \hline\hline

Spatial SVM   & 78.6\%          & 74.4\%                          \\ 
Motion SVM      & 78.2\%          & 57.9\%                          \\ 
Audio SVM 		& 16.7\% 		& 22.1\%		\\
\hline\hline

SVM-EF            & 86.9\%          & 75.9\%                          \\ 
SVM-LF            & 85.4\%          & 75.1\%                          \\  
SVM-MKL           & 87.1\%          & 75.6\%                          \\ \hline  \hline
NN-EF             & 86.6\%          & 76.1\%                          \\ 
NN-LF             & 85.4\%          & 75.4\%                          \\ 
M-DBM             & 87.0\%          & 76.0\%                          \\ 
RDNN    & 88.4\%        & 76.2\%              \\ \hline \hline

Non-regularized Fusion Network          
          & 87.2\%      & 75.8\%              \\ 
Regularized Fusion Network
          & {\bf 88.7\% }   & {\bf 76.7\% }             \\ \hline
\end{tabular}
\end{center}
\end{table}

\subsubsection{Feature Fusion}
We now move on to evaluate the proposed regularized feature fusion network and compare with competing methods. Table~\ref{tbl:fusion} presents the results and comparisons. In particular, the first group compares the results of the spatial, motion and audio features using SVMs. This set of experiments serves as baselines to better understand the improvements of fusion using SVM classifiers (summarized in the second group of~\ref{tbl:fusion}). See Table~\ref{tbl:lstm} for results that are directly obtained from CNN models. We also compare with alternative neural network based fusion methods as summarized in the third group in Table~\ref{tbl:fusion}.
Finally, we report the results of our method in the bottom row.

From the table, we can make the following observations (1) the fusion of multiple features offers performance gains on both UCF-101 and CCV and the improvements on UCF-101 are more significant than those on CCV; (2) the proposed feature fusion approach outperforms other neural network based methods; (3) the performance gain over the regularizer-free M-DBM network confirms modeling feature relationships is important during fusion; (4) our framework also outperforms RDNN slightly at a much lower cost as aforementioned.

To evaluate the contribution of norms in the objective function, we perform an ablation study and report the performance of the same network without any regularizers. Compared with the full model, the performance of the regularizer-free network drops 1.5\% on UCF-101 and 0.9\% on CCV.


\subsubsection{The Hybrid Framework}
We now discuss the effectiveness of the entire hybrid deep learning framework. In particular, we linearly average classification scores computed from the two LSTM models and the feature fusion network, which offers promising results, a mean accuracy of 92.1\% on UCF-101 and an mAP of 84.0\% on CCV (shown in Table~\ref{tbl:comparison}), outperforming alternative methods by clear margins. The entire hybrid framework improves 3.4 and 7.3 percentage points over the regularized fusion network (in Table~\ref{tbl:fusion}) on UCF-101 and CCV respectively, which stems from the combination with temporal clues captured by LSTM models. It is worth noting that our framework also achieves better performance than simple late fusion method (last row in Table~\ref{tbl:lstm}), which performs fusion with the same set of features.
 
\begin{figure}[t!]
\centering
\epsfig{file=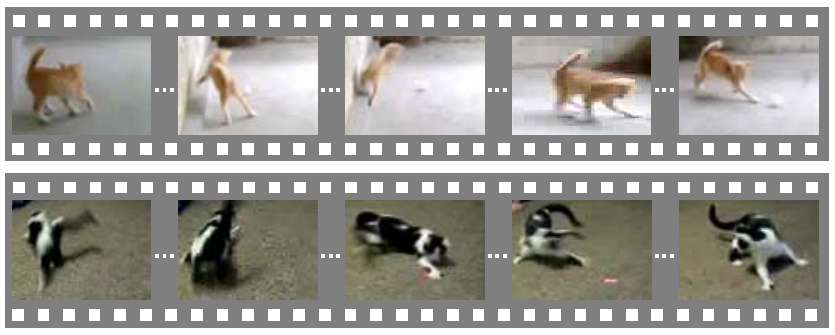, scale=1}
\caption{Two example videos of class ``cat" in the CCV dataset with similar temporal clues over time.}
\label{fig:example}
\end{figure}

For categories like ``graduation'' and ``birthday'' party in CCV, it is easy to understand that the fusion with temporal clues could assist recognition. We also examine other categories like ``cat" and ``dog" to see if there are certain temporal patterns. Interestingly, as illustrated in Fig~\ref{fig:example}, we found many ``cat" videos depicting a cat chasing objects or laser on the floor. Though the temporal order is not explicit, it could be captured by LSTM model for improved performance.



\begin{figure}[t!]
\centering
\includegraphics[width=0.7\linewidth]{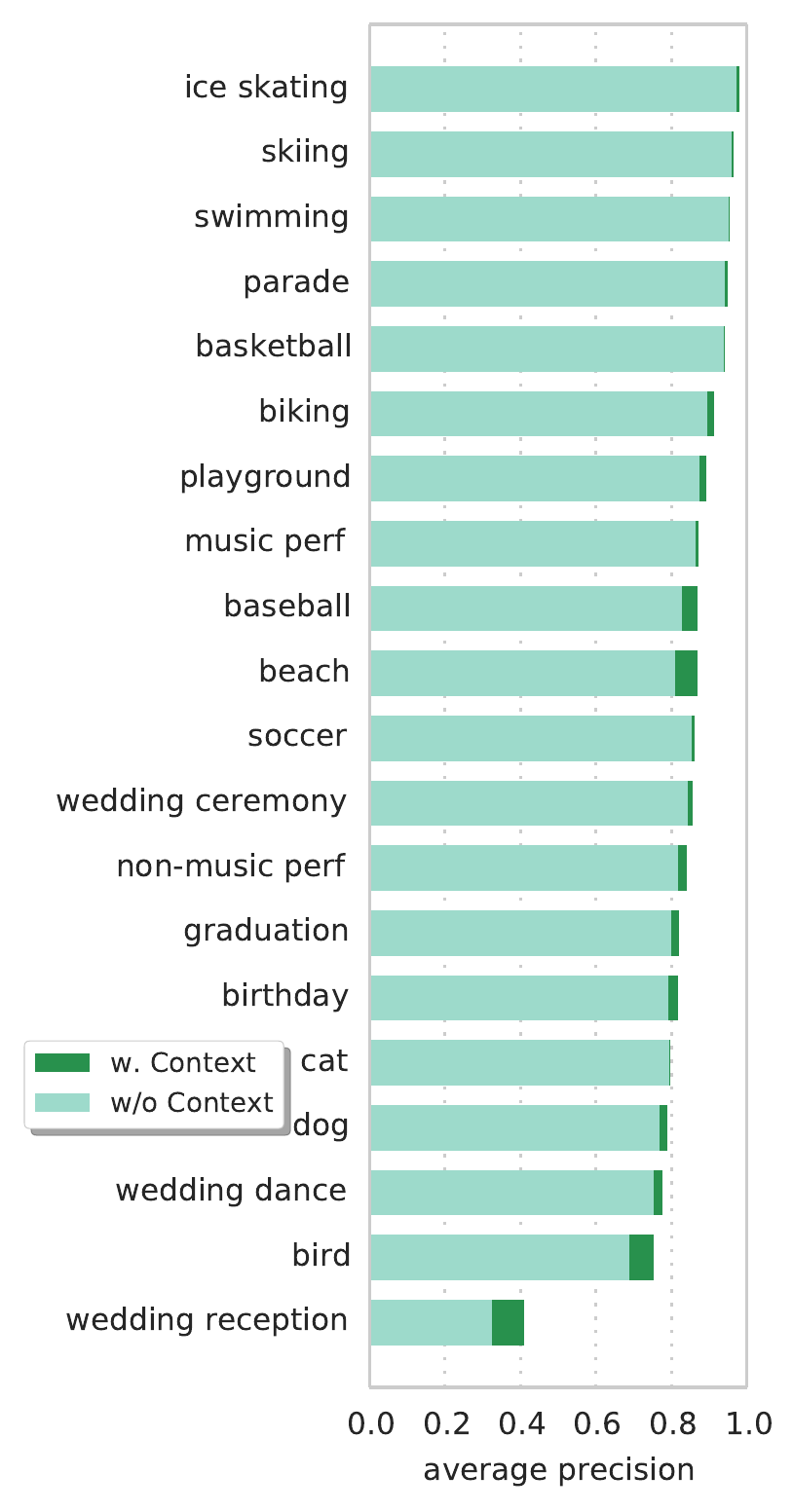}
\caption{Per-class average precision with and without contextual refinement on CCV. }
\label{fig:perclassCCV}
\end{figure}

Finally, we refine the prediction scores from the hybrid framework using semantics context. The result are summarized in the last row of~Table~\ref{tbl:comparison}. The contextual refinement is easy to perform but very effective, offering 1.0\% and 0.5\% performance gain over the original prediction scores. This confirms our assumption that related classes can assist the recognition of a class of interest.  {\color{black}In addition, we also compare with DASD~\cite{iccv09:dasd}, which utilizes context in a graph diffusion framework. Our context modeling method outperforms DASD by 0.7 and 0.3 percentage points with much lower computational complexity on UCF-101 and CCV, respectively. 

We further demonstrate per-class average precision after contextual refinement on CCV in Figure~\ref{fig:perclassCCV}. As can be seen from the figure, contextual refinement improves over the original model for nearly all classes. In addition, for classes with lower performance like ``bird'' and ``wedding reception'', the performance gains are more significant, resulting from the useful information borrowed from related classes.}

\subsubsection{Speed Efficiency}. To investigate the efficiency of our framework, we report the average time to classify a UCF-101 video clip using a single NVIDIA Telsa K40 GPU once the network is trained. Given a video clip, it takes around 4.5 seconds to compute RGB frames, optical flow images and audio spectrograms. The extraction of spatial, motion and audio CNN features takes 12 seconds. Finally, computing and refining the prediction scores from the LSTM and the feature fusion network can be finished in 4.3 seconds.

\begin{table}[t!]
\begin{center}
\caption{\label{tbl:comparison} Comparison with state-of-the-art results.}
\begin{tabular}{|c|c||c|c|} 
\hline
\multicolumn{2}{|c||}{UCF-101}                &    \multicolumn{2}{c|}{CCV}        \\ \hline \hline

Donahue \etal~\cite{Donahue2015}  & 82.9\%         & Lai \etal~\cite{Lai2014} & 43.6\%\\ 

Srivastava \etal~\cite{Srivastava2015} 
                                             & 84.3\%         & Jiang \etal~\cite{icmr11:consumervideo} & 59.5\% \\
                                             
Wang \etal~\cite{wang2013action}                        & 85.9\%         & Xu \etal~\cite{xu2013feature}    & 60.3\% \\ 
Tran \etal~\cite{Tran2015}                         & 86.7\%        & Ma \etal~\cite{DBLP:journals/ijcv/MaY14}   & 63.4\% \\ 
Simonyan \etal~\cite{DBLP:conf/nips/SimonyanZ14}            & 88.0\%         & Jhuo \etal~\cite{MVA:audiovisual} & 64.0\%\ \\ 

Ng \etal~\cite{Ng2015}                       & 88.6\%      & Ye \etal~\cite{ye2012robust}      & 64.0\% \\ 
Lan \etal~\cite{lan2014beyond}                        & 89.1\%         & Liu \etal~\cite{liu2013sample}    & 68.2\% \\ 
Zha \etal~\cite{Zha2015}                    & 89.6\%         & Wu \etal~\cite{Wu2015}   & 83.5\% \\ 
Wang \etal~\cite{Wang2015}                        & 91.5\%         & Nagel \etal~\cite{Nagel2015}           &  71.7\% \\ 
Wang \etal~\cite{Wang2016}			& 92.4\%		& & \\ \hline \hline

Hybrid Framework  &  92.1\% & Hybrid Framework &  84.0\% \\ \hline
Hybrid Framework-DASD &  92.4\% & Hybrid Framework-DASD &  84.2\% \\ \hline
Contextual Refinement  &  {\bf 93.1}\% & Contextual Refinement &  {\bf 84.5}\% \\ \hline

\end{tabular}
\label{tb:comparison}
\end{center}
\vspace{-0.2in}
\end{table}

\subsubsection{Comparison with State of the Arts}
We also compare with several state-of-the-art results on both datasets. Results are summarized in in Table~\ref{tbl:comparison}. We can see from the table that the proposed hybrid deep learning framework produces strong performance on both datasets. Different from works that obtain competitive results on UCF-101 using dense trajectory features~\cite{wang2013action,Zha2015}, our framework is built upon neural networks with an aim to learn feature representations. Our proposed approach improves the original twos stream CNN by incorporating temporal and audio modeling as well as better fusion methods. Notice that a few recent approaches also leverage temporal information with LSTMs~\cite{Donahue2015,Srivastava2015}; they utilized different CNN models to compute features, and hence the results are not directly comparable. Notice that we expect further performance improvements with more advanced neural networks like ResNet on UCF-101~\cite{Wang2016a,Feichtenhofer2016}.
On the CCV dataset, the proposed framework outperforms all the recent approaches that are designed to perform fusion by clear margins~\cite{xu2013feature,ye2012robust,MVA:audiovisual,DBLP:journals/ijcv/MaY14,liu2013sample,mm14:videoclassification}.

\section{Conclusions}
\label{sec:conclusion}
In this paper, we have proposed a novel hybrid deep learning framework to integrate a comprehensive set of multimodal clues for video categorization. More specifically, we utilize three independent CNN models operating on static frames, stacked optical flow images and audio spectrograms to compute spatial, motion and audio features, respectively. In order to utilize the long-range temporal clues in videos, we apply two LSTM models with the spatial and motion features as inputs. Since different features characterize the same video from different perspectives, we employ a regularized feature fusion network that derives a unified feature representation for recognizing video semantics. Finally, we also refine the classification scores, the linear combination of LSTM models and feature fusion network, with semantic contextual relationships. 

Through an extensive set of experiments on two challenging benchmarks, we demonstrate that (1) the LSTMs, modeling the long-range temporal information in video sequences through an explicitly recurrent manner, are highly complementary with CNNs; (2) the rich contextual relationships among video semantics in a simple yet effective way to further refine predictions for improved performance. The experimental results provide strong quantitative evidence that our framework achieves promising results, outperforming competing methods with clear margins.

\section*{Acknowledgment}
This work was supported in part by two grants from NSF China (\#61622204, \#61572134) and two grants from STCSM, Shanghai, China (\#16QA1400500, \#16JC1420401).

\scriptsize
\bibliographystyle{abbrv}
\bibliography{reference}


\end{document}